# Iterative decoding of Generalized Parallel Concatenated Block codes using cyclic permutations

Hamid Allouch[1], Idriss Chana[2] and Mostafa Belkasmi[3]

[1] University Mohamed V SOUISSI , ENSIAS
Rabat, 10000, MOROCCO
hamid.allouch@gmail.com

[2] University Mohamed V SOUISSI, ENSIAS
Rabat, 10000, MOROCCO
id_chana@yahoo.fr

[3] University Mohamed V SOUISSI, ENSIAS
Rabat, 10000, MOROCCO
belkasmi@ensias.ma

**Abstract**

Iterative decoding techniques have gain popularity due to their performance and their application in most communications systems. In this paper, we present a new application of our iterative decoder on the GPCB (Generalized Parallel Concatenated Block codes) which uses cyclic permutations. We introduce a new variant of the component decoder.
After extensive simulation; the obtained result is very promising compared with several existing methods. We evaluate the effects of various parameters component codes, interleaver size, block size, and the number of iterations.
Three interesting results are obtained; the first one is that the performances in terms of BER (Bit Error Rate) of the new constituent decoder are relatively similar to that of original one. Secondly our turbo decoding outperforms another turbo decoder for some linear block codes. Thirdly the proposed iterative decoding of GPCB-BCH (75, 51) is about 2.1dB from its Shannon limit.

## 1. Introduction

Since the initial proposal of 'Turbo Codes' by Berrou et al in 1993 [1], the iterative principle has been extended to other new codes families [2, 3, 4, 7]. In [5], we have developed a new soft decoding algorithm (SIHO decoder) based on cyclic propriety for the most linear block codes. We use the cyclic permutation which stabilizes all cyclic codes and a set of test vectors obtained by flipping least reliable bits of the received codeword, the algorithm is considered as a light soft version of permutation decoding.
In [6] we focus on changing SIHO decoder presented in [5] in order to reduce its complexity by studying some parameters. We compute its soft output using extrinsic information according to Soleymani et al [7]. We realized a turbo decoding, by adopting the Pyndiah's connection scheme without the famous coefficients $\alpha$ and $\beta$. Both versions in [5] and [6] can be applied for any cyclic codes, particularly QR (Quadratic Residue), BCH and DSC (Difference Set cyclic) codes.
In this work we present:
The version developed in [6] is an improvement of these in [5] in terms of complexity.
In this perspective and trough this work, we propose a new variant of the algorithm presented in [6] in order to reduce more its complexity and after, to used it as a component decoder within a turbo decoding of General parallel concatenated block codes (GPCB). In the proposed turbo decoder designed to decode iteratively the the GPCB codes, we adopt the Soleymani's [7] soft output to generate the extrinsic information and the Pyndiah's connection scheme.
We demonstrate the applicability of our efficient algorithm on new families of codes i.e. the GPCB codes.
The performances of this new iterative decoder are investigated using simulations. The effects of various component codes, the number of iterations, block size; interleaver size and pattern are studied.
Iterative decoding of concatenated codes is a way of using long powerful codes while keeping the new variant decoder relatively efficient. The concept of our turbo decoding can be applied to any concatenated codes constructed from cyclic codes. In this paper, we evaluate our new decoder on the families of generalized parallel-concatenated block (GPCB) codes.
After an overview of the decoder algorithm variant adopted here, and its complexity, we show significant reducing of the number of useful cyclic permutations and test sequences. On other hand we have obtained three interesting results by analyzing the simulations carried out and by comparing with other works.

The rest of this paper is organized as follow, in section II we present the component decoder, section III describe the soft output structure of the component decoder. Section IV describes GPCB codes construction and their iterative decoding. The simulation results are analyzed in section V. Section VI concludes this paper.

## 2. Component Decoder

In the previous algorithm [5] the authors use two loops, the first is using n circular permutations for received word, and the second one processes $2^p - 1$ error patterns generated by the parameter $p$.

### 2.1 The description of improvement of our component decoder

In [6] they restrict the number of permutations in order to reduce its complexity considering only $k$ cyclic permutations; the target permutation is taken according to the maximum sum of the soft values within $k$ positions of the received word R.

In this paper, we present a light version of this decoder algorithm by including a stopping criteria "Threshold" introducing the confidence value concept.

Figure 1 describes the light version, in which we continue to generate the error pattern and encoding systematic part until we reached the threshold value (such $E_i^j <$ Threshold).

We stopped the generation of test sequences and goes out of all the rest of permutations, if the threshold is reached.

We continue to generate test sequences with $k$ cyclic permutations if threshold is not reached.

The input of the decoder, when the channel perturbed by a white Gaussian noise, is equal to $r_c = u + b_r$, where $r_c = (r_1 \ldots r_j \ldots r_{n_i})$ is the observed vector, $u = (u_1 \ldots u_j \ldots u_{n_i})$ $u_j = \pm 1$ the emitted codeword and $b_c = (b_1 \ldots b_j \ldots b_{n_i})$ is the white noise whose components $b_j$ have zero average and variance $\sigma^2$.

The received word $r_c$ is sorted in descending order and $\sigma$ denotes the permutation associated to this sort: $q = \sigma(r_c)$

$$\begin{cases} q_i = r_{c(pi)} & 0 \leq pi \leq \text{n-1} \\ |q_0| \geq |q_1| \ldots \geq |q_{n-1}| \end{cases} \quad (1)$$

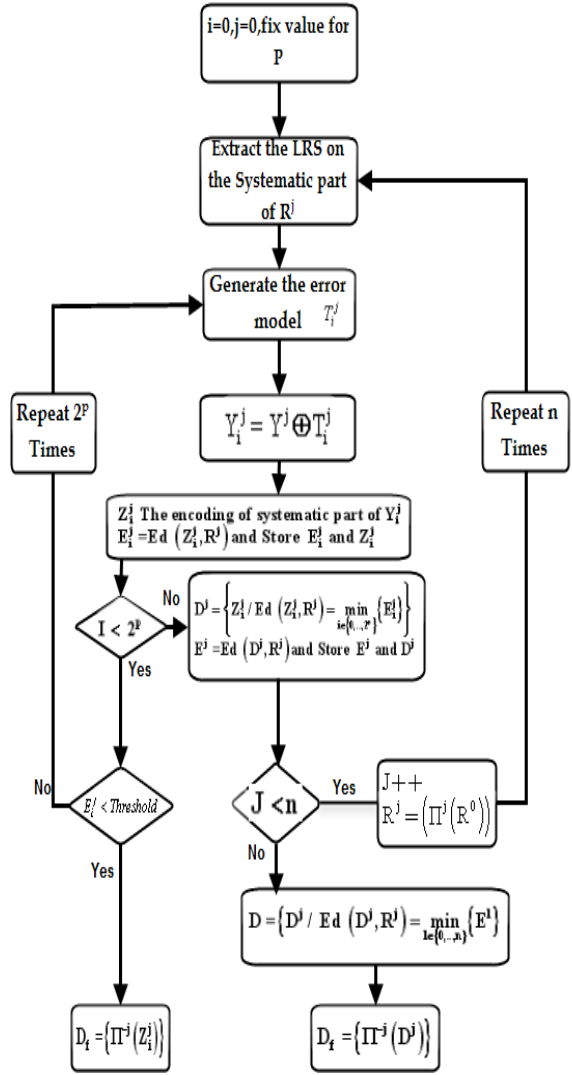

Fig. 1 The flow chart of the algorithms with stopping criteria

### 2.2 Implementation and simulations results

The new algorithm is very attractive because it need not swap over $2^p$ extra test vector and $k$ permutations. In addition, it's very simple to implement and has a low complexity. Using the threshold for comparison, we show in figure 2 and 3 that the complexity is reduced significantly by increasing the SNR. Precisely in figure 3 at the SNR=6 we have 98% of test sequence gain complexity, by using the proposed algorithm.

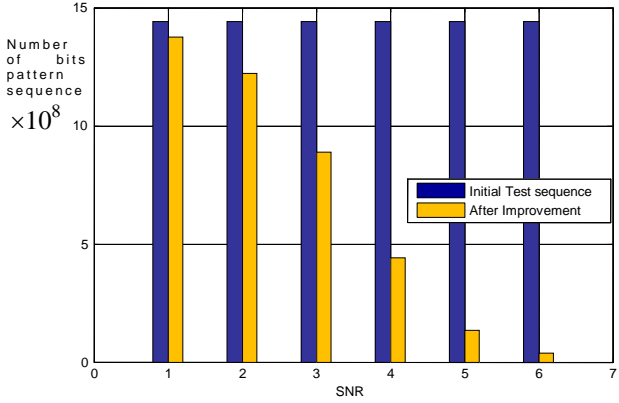

Fig. 2   Performances Comparison by varing SNR.

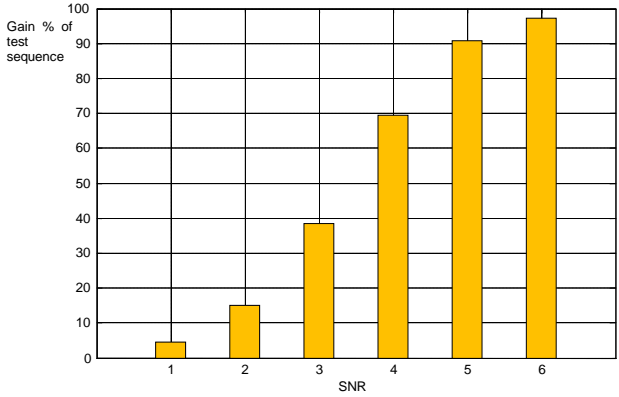

Fig. 3   Gain pourcent of our algorithm by varing SNR

We show in figure 4 that the Performances of our light version in terms of BER compared with [5] are the same.

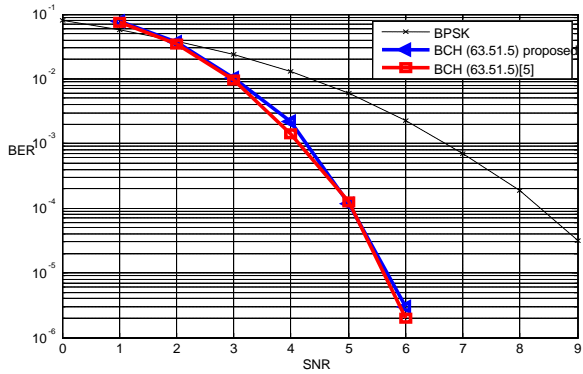

Fig. 4   Performances Comparison using different permutations for BCH (63, 51,5).

But the computational complexity of our variant algorithm is less complex than the [6] and [7] as show in table I.
Table I below present the analysis computation of complexity of three algorithms.

For BCH (63, 51, 5) and at SNR=5 we have $K_a \approx 51 = k$.

Table 1: Computational of Complexity for the Three Variants

| Algorithm | Complexity |
|---|---|
| SIHO Variant in [5] | $O\left(2^P n.\log(n)(n+\log(n-k))\right)$ |
| SIHO Variant in [6] | $O\left(2^{P'+1} k.\log(n)(n+\log(n-k))\right)$ |
| Our Algorithm | $O(K_a.\log(n)(n+\log(n-k)))$ <br> $K_a \prec\prec 2^P.n$ |

## 3. Confidence value and Soleymani Soft Output

3.1 Confidence value:

The concept of confidence value denoted by $\Phi$ is detailed more in [7]. In this subsection, we will content ourselves with give a brief description.

Let $X = \{x_0, x_1, ..., x_{n-1}\}$ be the transmitted codeword; $\Phi = P\{D = X \mid R\}$ be the probability that the decoder makes correct decision $D = \{d_0, d_1, ..., d_{n-1}\}$ giving received sequence $R = \{r_0, r_1, ..., r_{n-1}\}$. In other words, it is the evaluation of the decoder's decisions. Computing $\Phi$ is impossible for a practical implementation, thus estimation has to be performed. To estimate $\Phi$, Soleymani et al. [7] adopted a distance destructive denoted by $Dist_{dest}$ as a metrics measure between $R$ and $D$ where only the positions increasing the Euclidian distance contribute, that is to say, where the noise vector R has a different polarity from the decision vector $D$. The following formula (2) illustrates this.

$$Dist_{dest} = \sum_{j \in \text{DES}} \left(r_j - d_j\right)^2 \quad (2)$$

where $\text{DES} = \left\{ j \mid \left(r_j - d_j\right).d_j < 0 \right\}$

As we may notice, there is a relationship between the confidence value $\Phi$ and the destructive Euclidean distance $Dist_{dest}$. Using simulation software according to

Soleymani et al. [2], the influence of the variables Eb/No, P and the number of iterations may be omitted, and a confidence value $\Phi$ is treated as a function of destructive Euclidean distance, it can be written as:

$$\Phi = f\left(Dist_{dest}\right) \quad (3)$$

3.2 Computation of Soleymani's soft output

In this subsection, we are giving the computation of Soleymani's soft output as it is described in [7] that $X = \{x_0, x_1, ..., x_{n-1}\}$ is the transmitted code word; the symbol $x_j$ $j \in \{0, ..., n-1\}$ has certain confidence value $\Phi$. The probability of $x_j$ can expressed as:

$$P(x_j = \pm 1 \mid R) = P(x_j = \pm 1, D = X \mid R) + P(x_j = \pm 1, D \neq X \mid R) \quad (4)$$

The first term of (4) represents the probability value when the decoder gives a correct codeword. Applying Bayes' rule to this term will yield:

$$P(x_j = \pm 1, D = X \mid R) = P(x_j = \pm 1, D = X, R).P(D = X, R)$$
$$= P(x_j = \pm 1, D = X, R).\Phi \quad (5)$$

Since the decision bit $d_j$ is known, then

$$P(x_j = \pm 1, D = X \mid R) = \begin{cases} \phi & \text{if } d_j = x_j \\ 0 & \text{if } d_j \neq x_j \end{cases} \quad (6)$$

The second term in (4) represents the probability value when the decoder decides in favor of a wrong codeword. In this case, we consider the transmitted symbol $x_j$ corrupted by Gaussian noise. Thus:

$$P(x_j = 1, D \neq X) = \frac{\exp\left(\pm 2r_j / \sigma^2\right)}{1 + \exp\left(\pm 2r_j / \sigma^2\right)} \quad (7)$$

Once again, we apply Bayes's rule to the second term in (4) and we get

$$P(x_j = \pm 1, D \neq X \mid R) = P(x_j = \pm 1, D \neq X, R) . P(D \neq X \mid R)$$
$$= \frac{\exp\left(\pm 2r_j / \sigma^2\right)}{1 + \exp\left(\pm 2r_j / \sigma^2\right)} . (1 - \Phi) \quad (8)$$

Combining (4)-(8), the a posteriori probability of $x_j$ becomes

$$P(x_j = +1 \mid R) = \begin{cases} \phi + \dfrac{\exp\left(+2r_j/\sigma^2\right)}{1 + \exp\left(+2r_j/\sigma^2\right)} . (1-\Phi) & \text{if } d_j = +1 \\ \dfrac{\exp\left(+2r_j/\sigma^2\right)}{1 + \exp\left(+2r_j/\sigma^2\right)} . (1-\Phi) & \text{if } d_j = -1 \end{cases} \quad (9)$$

And

$$P(x_j = -1 \mid R) = \begin{cases} \dfrac{\exp\left(-2r_j/\sigma^2\right)}{1 + \exp\left(-2r_j/\sigma^2\right)} . (1-\Phi) & \text{if } d_j = +1 \\ \phi + \dfrac{\exp\left(-2r_j/\sigma^2\right)}{1 + \exp\left(-2r_j/\sigma^2\right)} . (1-\Phi) & \text{if } d_j = -1 \end{cases} \quad (10)$$

As in the traditional algorithm presented by Pyndiah in [9], we can obtain the extrinsic information $\omega_j$ by the following equation

$$\omega_j = \frac{\sigma^2}{2} \ln\left(\frac{P(x_j = +1 \mid R)}{P(x_j = -1 \mid R)}\right) - r_j \quad (11)$$

(11)

Substituting P (xj = +1 | R) and P (xj = −1 | R) from (6) and (7), we get

$$\omega_j = d_j \left(\frac{\sigma^2}{2} \ln\left(\frac{\Phi + \exp\left(2r_j d_j / \sigma^2\right)}{1 - \Phi}\right) - r_j d_j\right) \quad (12)$$

Unlike other list-based algorithms, soft outputs generated by previous formula (12) can directly injected into the next decoding stage without scaling by a weighting factor α.

## 4. GPCB codes and their iterative decoding

4.1 GPCB codes construction

The GPCB codes based on cyclic code as a form of concatenation are presented in figure 5; For decoding generalized parallel-concatenated block (GPCB) codes, we are using as component codes with an interleaver placed before the second block encoder. A block of $N = M * k$ data bits at the input of the encoder is subdivided to M sub-blocks each of k bits. Each k bits vector is encoded in order to produce n bits codeword. The input block is scrambled by the interleaver-denoted by Π- before entering the second encoder. The codeword of GPCB code consists of the input block followed by the parity check bits of both encoders.

In this contribution several interleaving techniques were invoked such as helical, random, block, diagonal, and cyclic interleaver.

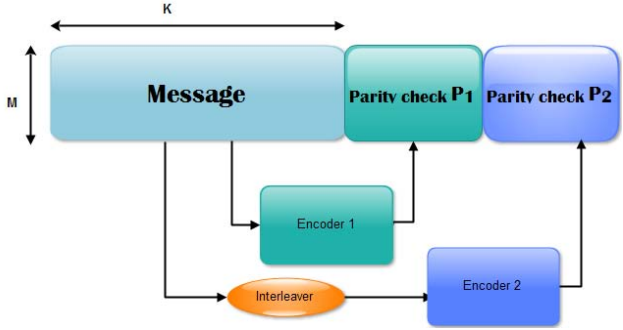

Fig. 5 The M.K Information bits are encoded twice by supplying the original information and its interleaved version.

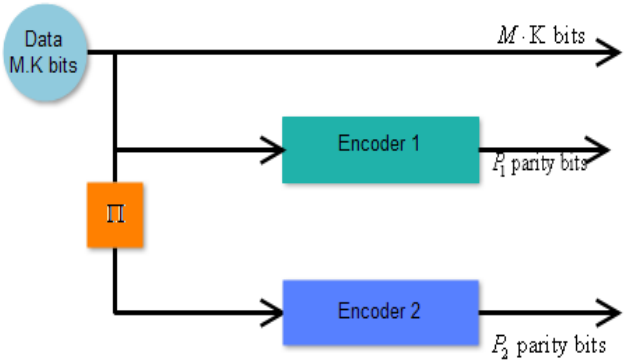

Fig. 6 Encoder structure of GPCB codes

A systematic GPCB code is based on two component systematic block codes, C1 with parameters (n1, k) and C2 with parameters (n2, k). Viewing the coding scheme of figure as single GPCB encoder, the length of the information-word to be encoded by the GPCB code is given by the size of the interleaver $N = M \times k$.

The first encoder produces $P_1 = M \times (n_1 - k)$ parity check bits. The second encoder produces $P_2 = M \times (n_2 - k)$ parity check bits. Thus the total number of parity bits generated by the GPCB encoder is $P = P_1 + P_2 = M \times (n_1 + n_2 - 2 \times k)$.

The length of the GPCB codeword is given by: $L = N + P = M \times (n_1 + n_2 - k)$. Consequently, the code rate of the PCB codes can be computed by:

$$R = \frac{N}{L} = \frac{k}{n_1 + n_2 - k}.$$

This implies that the GPCB code rate is independent of the interleaver size N. Below we give some examples of codes based on this construction.

Table 2: Some examples of PCB codes

| Component code | M | GPCB code |
|---|---|---|
| BCH(63, 51, 7) | 1 | GPCB-BCH(75, 51) |
| | 10 | GPCB-BCH(750, 510) |
| | 100 | GPCB-BCH(7500, 5100) |
| | 200 | GPCB-BCH(14000, 10200) |
| RQ(47, 24, 11) | 1 | GPCB-RQ(70, 24) |
| | 10 | GPCB-RQ(700, 240) |
| | 100 | GPCB-RQ(7000, 2400) |
| | 200 | GPCB-RQ(14000, 4800) |
| BCH(127, 106, 7) | 1 | GPCB-BCH(148, 106) |
| | 10 | GPCB-BCH(1480, 1060) |
| | 100 | GPCB-BCH(14800, 10600) |
| | 200 | GPCB-BCH(29600, 21200) |
| BCH(255, 215, 11) | 1 | GPCB-BCH(295, 215) |
| | 10 | GPCB-BCH(2950, 2150) |
| | 100 | GPCB-BCH(29500, 21500) |
| | 200 | GPCB-BCH(59000, 43000) |

### 4.2 GPCB iterative decoding

The decoding of the GPCB codes is iterative (see figure 7). The first component decoder uses the systematic information and the first parity check bits in order to generate extrinsic information as in Pyndiah's connection scheme.

This extrinsic information is used to update the reliabilities of the systematic information, which will be interleaved and feed into the second decoder with the second parity check bits received from the channel.

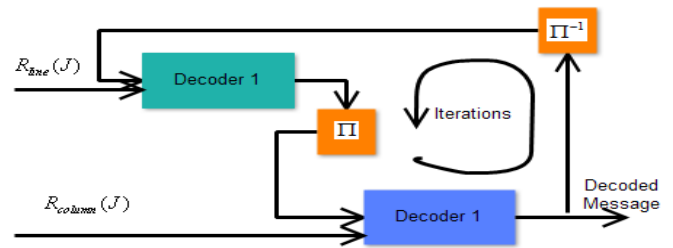

Fig. 7 The turbo decoder at Jth iteration

The second decoder also generates the extrinsic information, and then updates the reliabilities of the systematic information for the second time. The updated reliabilities will be des-interleaved and feed again into first decoder, for the next iteration. The process resume until a maximum number of iterations is reached.

# 5. Simulation results

In this section, the performance of the generalized parallel-concatenated BCH (GPCB-BCH) codes is evaluated. The transmission is over the additive white Gaussian noise (AWGN) channel and binary antipodal modulation. We are interested in the information bit error rate (BER) for different signal to noise ratios per information bit (Eb/N0) in dB. There are many parameters, which affect the performance of GPCB codes when decoded with iterative decoder.

Here we study the effect of the following parameters on the decoder as in table III, precisely the number of decoding iterations, the component codes, interleaver size and patterns

Table 1: The parameters of communication system

| Parameter name | Value |
|---|---|
| Modulation | BPSK |
| Channel | AWGN |
| Interleaver pattern | Semi-Random interleaver ; Diagonal interleaver; Cyclic interleaver; Block interleaver |
| Component decoder | Soleymani output and Pyndiah connection schema |
| Iterations | 1 to 10 (6 as default value) |
| Interleaver size | 1xk, 10xk, 50xk ,100xk, 200xk |

## 5.1 The Turbo effect (number of iterations)

The simulation results plotted in figure 8 show the performance of the iterative decoder for the GPCB-BCH (3750, 2550).

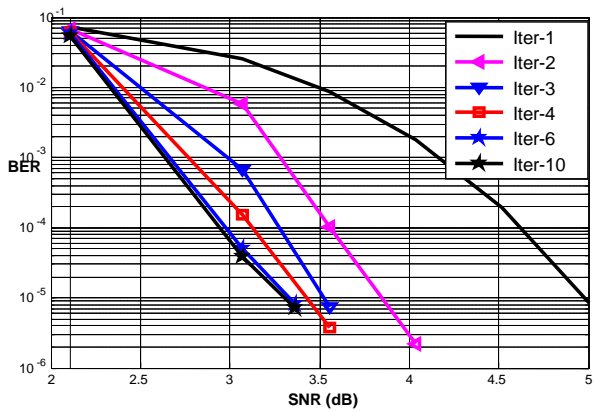

Fig. 8 Effect of iterations on Iterative decoding for G*PCB-BCH(3750,2550)* code

Figure 8 show that the slope of curves and coding gain are improved by increasing the number of iterations. At 10-5 about 1.8dB coding gain can be obtained after 4 iterations. After the 4th iteration, the amelioration of the coding gain becomes negligible because of the steep slope of the BER curve. The turbo phenomenon is well established. In the rest of this paper, the curves are done with 4 iterations.

## 5.2 The effect of block size (M parameter)

Footnotes should be typed in singled-line spacing at the bottom of the page and column where it is cited. Footnotes should be rare.

Figure 9 shows the BER versus SNB results of the BCH (75, 51) code with M varying from 1 to 200. By increasing M from 1 to 100, about 3.0dB coding gain can be obtained at 10-5 and little gain can be obtained by further increasing the parameter M.

Now we consider the QR (70, 24) code. The performance is shown in figure 10. For this code, the coding gain increase with M. At 10-5 about 3.0dB coding gain can be obtained by increasing M from 1 to 100. The amelioration becomes inconsiderable while the parameter M is greater or equal to 200.

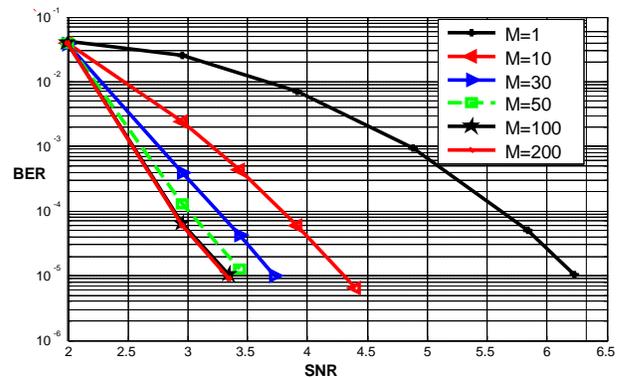

Fig. 9 Effect of parameter M on Iterative decoding on BCH(63,51)

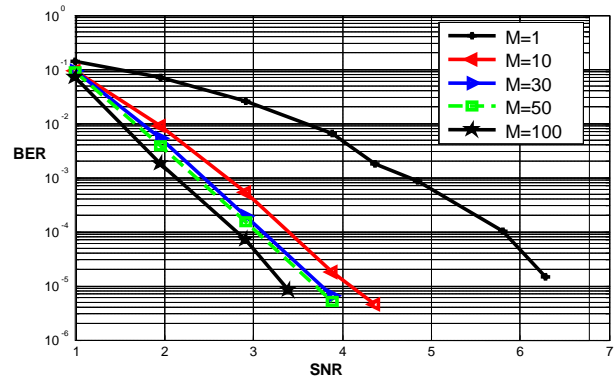

Fig. 10 Effect of parameter M on Iterative decoding of *QR(47, 24)* code

## 5.3 GPCB iterative decoding The effect of interleaver

To study the influence of the interleaver pattern on the GPCB codes performance, we have evaluated the BER of the GPCB-BCH (750,510) code using different interleavers.

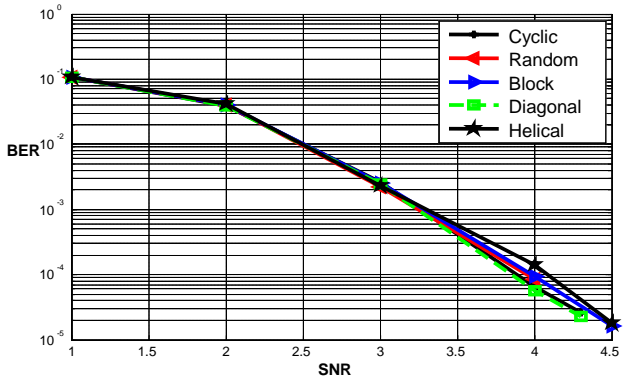

Fig. 11 Interleaver structure effect on G*PCB-BCH(750, 510)* code

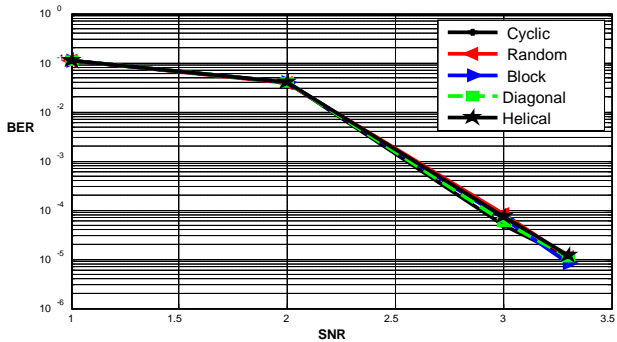

Fig. 12 Interleaver structure effect on Iterative decoding for G*PCB-BCH(7500, 5100)* code

Such as diagonal, cyclic, block and random interleaver with two different value of parameter M (M=10, M=100). The figure 11 shows the results For M=10 and figure 12 for M=100, we observe that the cyclic and diagonal are little good than random, block and helical ones for m=10. For M=100, we observe that all these interleavers are similar.

## 5.4 Comparisons with other works and Shannon limits.

To evaluate the performance of the parallel-concatenated block codes, we compare the coding gain at 4th iteration of the codes GPCB-BCH (7500, 5100).

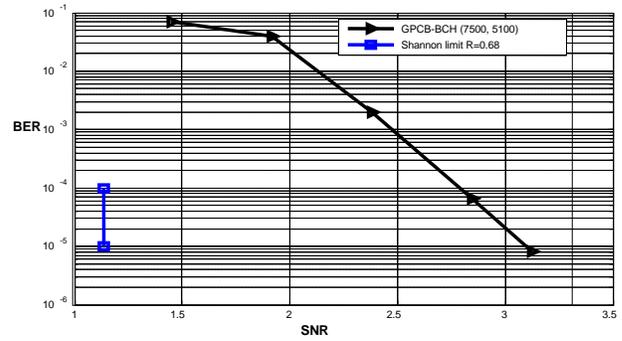

Fig. 13 Performance comparison of GPCB-BCH (75, 51) code and its position from Shannon limit

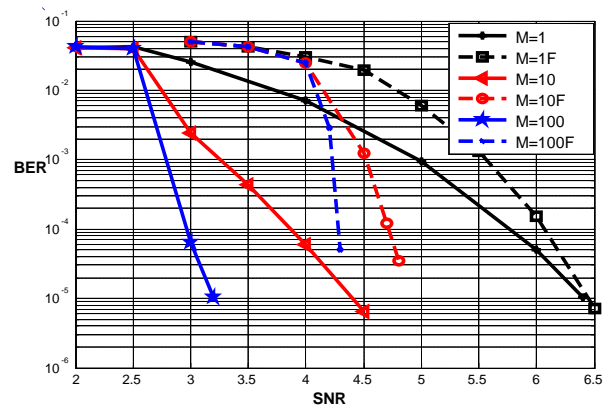

Fig. 14 Performance comparison between proposed GPCB codes and Farchane et al. GPCB (dotted line), at 6 iterations.

In figure 13, we observe that performance of this code about 2.1 dB away from their Shannon Limit.
In the other hand the comparison between our decoder and the algorithm of Farchane et al. [10], for the same rate in figure 14, we shows that our decoder outperforms the Farchane et al. algorithm; for GPCB-BCH (7500, 5100) code and at 6 iterations, for BER=10-5 our decoder gain is about 1.3 dB than [10].

## 6. Conclusions

In this paper, we have demonstrated the application of the improved and efficient SIHO decoder algorithm as a component decoder in a turbo decoder to decode the GPCB codes. We have investigated the effects of various component codes, the number of iterations, interleaver size and pattern using simulations. The results show that the slope of curves and coding gain are improved by increasing the number of iterations and/or the interleaver size (the parameter M).

From the simulations, we retain that GPCB codes based on GPCB-BCH (75, 51) is about 2.1dB from Shannon limit. The decoder presented by Farchane et al [10] can be applied only for codes with algebraic component decoder; however our decoder is practical for all GPCB codes based on cyclic codes. The obtained results look very promising and open new perspectives. The extension of this study is to investigate this decoding for serial concatenated block codes and for the family of codes RS.

**H.Allouch** Received the Telecommunication & Informatics engineering degree from INPT (Institute National of Posts and Telecommunications), Rabat, Morocco in 2003. Currently he is preparing his PhD degree in Computer Science and Engineering at ENSIAS, Rabat, Morocco. His current research domains are Information and Coding Theory, mobile 3G/4G network security, wireless communications, and NGN interworking and QoS.

**I.Chana** Received his license in Computer Science and Engineering in Jun-2001 and Master in Computer Science and telecommunication from University of Mohammed V - Agdal, Rabat, Morocco in 2006. Currently he is doing his PhD in Computer Science and Engineering at ENSIAS (Ecole National d'Informatique et Anlyse Systme), Rabat, Morocco. Her areas of interest are Information and Coding Theory.

**M. Belkasmi** Is recently a PES professor at ENSIAS (Ecole National d'Informatique et Analyse des Systemes); head of Telecom and Embedded Systems Team at SIME Lab .He had PhD at Paul Sabatier (Toulouse III) University in 1992. His actual research domains are mobile and wireless communications, 3G/4G networks, QoS NGN, and Information and Coding Theory.